
\input harvmac
\def\np{Nucl. Phys. }
\def\pl{Phys. Lett. }

\def\prl{Phys. Rev. Lett. }

\def \CL{{\cal L}}

\def \CK{{\cal K}}

\def\frac#1#2{{#1\over#2}}

\def\half{\frac12}

\def\journal#1&#2(#3){\unskip, \sl #1\ \bf #2 \rm(19#3) }
\def\andjournal#1&#2(#3){\sl #1~\bf #2 \rm (19#3) }

\lref\cern{J. Ellis, N.E. Mavromatos and D.V. Nanopoulos, ``On the
Evaporation of Black Holes in String Theory,'' CERN-TH
6309/91 ACT-53 CTP-TAMU-90/91.}
\lref\ceqone{E. Brezin, V. Kazakov, and Al. Zamolodchikov, Nucl. Phys.
{\bf B338}
(1990) 673; D. Gross and N. Miljkovi\'c, Phys. Lett {\bf 238B} (1990) 217;
P. Ginsparg and J. Zinn-Justin, Phys. Lett {\bf 240B} (1990) 333;
G. Parisi, Phys. Lett. {\bf 238B} (1990) 209.}
\lref\wadia{G. Mandal, A.M. Sengupta and S.R. Wadia,
Mod. Phys. Lett. {\bf A6} (1991) 1465.}
\lref\dvv{R. Dijkgraaf, H. Verlinde and E. Verlinde, ``String
Propagation in a Black Hole Geometry'' PUPT-1252,
IASSNS-HEP-91/22.}
\lref\gk{K. Gawedzki and A. Kupiainen, to appear; K. Gawedzki,
``Noncompact WZW Conformal Field Theories,''  IHES-P-91-73.}
\lref\lenny{L. Susskind, unpublished.}
\lref\marsha{E.J. Martinec and S.L. Shatashvili, ``Black Hole Physics
and Liouville Theory'' EFI-91-22.}
\lref\mat{J. Ambj{\o}rn, B. Durhuus, J. Fr\"ohlich, Nucl. Phys.
{\bf B257} (1985) 433; F. David, Nucl. Phys. {\bf B257} (1985) 45;
V. Kazakov, Phys. Lett. {\bf 150B} (1985) 282;
V. Kazakov, I. Kostov, A. Migdal, Phys. Lett. {\bf 157B} (1985) 295.}
\lref\dbsc{E. Br\'ezin and V. Kazakov, Phys. Lett. {\bf 236B} (1990) 144;
M. Douglas and S. Shenker, Nucl. Phys. {\bf B335} (1990) 635;
D. Gross and A. Migdal, Phys. Rev. Lett. {64} (1990) 127.}
\lref\cargese{For further references see
{\it Matrix Models, String Field Theory and
       Topology}, Proceedings of the Carg\`ese Workshop on Random
      Surfaces, Quantum Gravity and Strings 1990, O. Alvarez, E.
      Marinari, and P. Windey (eds.), Plenum, to appear.}
\lref\dewitt{B.S. deWitt, Phys. Rev. {\bf 160} (1967) 1113.}
\lref\phitime{F. David
and E. Guitter, Europhys.Lett. {\bf 3} (1987) 1169; \np {\bf B293}
(1988) 332; S.R. Das, S. Naik and S.R. Wadia, Mod. Phys. Lett.  {\bf A4}
(1989) 1033; J. Polchinski, Nucl. Phys. {\bf B324} (1989) 123; T. Banks
and J. Lykken, \np {B331} (1990) 173.}
\lref\natiliou{N. Seiberg, Notes on Quantum Liouville Theory and Quantum
Gravity.  In: {\it Common Trends in Mathematics and Quantum Field
Theory} Proc. of the 1990 Yukawa International Seminar, Kyoto.  Edited
by T.  Eguchi, T. Inami and T. Miwa.  To appear in the Proc. of the
Cargese meeting, Random Surfaces, Quantum Gravity and Strings, 1990.
RU--90--29.}
\lref\joeliou{J. Polchinski,``Remarks on the Liouville Field Theory,''
Texas preprint UTTG-19-90, Presented at Strings '90 Conf., College
Station, 1990.}
\lref\ctthrn{T.L. Curtright and C.B. Thorn, \prl {\bf 48} (1982) 1309;
E. Braaten, T. Curtright and C. Thorn, \pl {\bf 118B}
(1982) 115; Ann. Phys. {\bf 147} (1983) 365;
E. Braaten, T. Curtright, G. Ghandour and C. Thorn, \prl {\bf 51}
(1983) 19; Ann. Phys. {\bf 153} (1984) 147.}
\lref\gervais{J.-L. Gervais and A. Neveu, \np
{\bf 199} (1982) 59; {\bf B209} (1982) 125;{\bf B224} (1983) 329;
{\bf 238} (1984) 125; 396; \pl {\bf 151B} (1985) 271;
J.-L. Gervais, LPTENS 89/14; 90/4.}
\lref\bh{S. Elitzur, A. Forge and E. Rabinovici, \np {\bf B359} (1991)
581;  E. Witten, Phys. Rev. {\bf D44} (1991) 314;
G. Mandal, A. Sengupta and S. Wadia, Mod. Phys. Lett. {\bf A6} (1991)
1685;
M. Rocek, K. Schoutens and A. Sevrin, Phys. Lett. {\bf B265} (1991) 303.}
\lref\cw{see for example C. Callan and F. Wilczek, Nucl. Phys. {\bf
B340} (1990) 366.}
\lref\pol{A. Polyakov, Phys. Lett. {\bf 103B} (1981) 207, 211.}
\lref\plsltw{A. Polyakov, Mod. Phys. Lett. A {\bf 2} (1987) 893;
V. Knizhnik, A. Polyakov and A. Zamolodchikov, Mod. Phys. Lett.
{\bf A3} (1988) 819; F. David, Mod. Phys. Lett. {\bf A3} (1988)
1651; J. Distler and H. Kawai, Nucl. Phys. {\bf B321} (1989) 509.}
\lref\mss{G. Moore, N. Seiberg, and M. Staudacher, \np {\bf B362}
(1991) 665.}
\lref\mooreplesser{G. Moore, M. R. Plesser and S. Ramgoolam. ``Exact S
Matrix for 2D String Theory'' YCTP-P35-91.}
\lref\joecone{J. Polchinski, \np {\bf B346} (1990) 253.}
\lref\kutdf{P. Di Francesco and D. Kutasov, Phys. Lett. {\bf B261}
(1991) 385.}
\lref\joeclas{J. Polchinski, Nucl. Phys. {\bf B362} (1991) 125.}
\lref\moore{G. Moore,``Double--Scaled Field Theory at $c=1$,''
Rutgers preprint RU--91--12.}
\lref\dj{S. Das and A. Jevicki, Mod. Phys. Lett. {\bf A5} (1990) 1639;
K. Demeterfi, A. Jevicki, and J.P. Rodrigues,
Nucl. Phys.  {\bf B362} (1991) 173; {\bf B365} (1991) 499.}
\lref\ms{G. Moore and N. Seiberg,
``From Loops to Fields in $2D$ Quantum Gravity''
RU-91-29, YCTP-P19-91.}
\lref\grossmende{D. Gross and P. Mende,  Phys. Lett. {\bf 197B} (1987)
129; Nucl. Phys. {\bf B303} (1988) 407.}
\lref\gpy{D. Gross, M. Perry and L. Yaffe, Phys. Rev. {\bf D25} (1982)
330.}
\lref\hawk{S. W. Hawking in {\it General Relativity: An Einstein
Centennial Survey}, S. W. Hawking and W. Israel eds. (Cambridge
University Press, London 1979).}
\lref\curt{T. Curtright and G. Ghandour, \pl {\bf 136B} (1984) 50.}
\Title{\vbox{\baselineskip12pt\hbox{RU-91-53}}}
{{\vbox {\centerline{A Note on Background (In)dependence}
}}}

\centerline{\it Nathan Seiberg and Stephen Shenker}
\smallskip
\centerline{Department of Physics and Astronomy}
\centerline{Rutgers University, Piscataway, NJ 08855-0849, USA}
\centerline{seiberg@physics.rutgers.edu, shenker@physics.rutgers.edu}
\vskip .2in

\noindent
In general quantum systems there are two kinds of spacetime modes, those
that fluctuate and those that do not.  Fluctuating modes have
normalizable wavefunctions.  In the context of 2D gravity and
``non-critical'' string theory these are called macroscopic states. The
theory is independent of the initial Euclidean background values of
these modes. Non-fluctuating modes have non-normalizable wavefunctions
and correspond to microscopic states.  The theory depends on the
background value of these non-fluctuating modes, at least to all orders
in perturbation theory. They are superselection parameters and should
not be minimized over.  Such superselection parameters are well known in
field theory.  Examples in string theory include the couplings $t_k$
(including the cosmological constant) in the matrix models and the mass
of the two-dimensional Euclidean black hole.  We use our analysis to
argue for the finiteness of the string perturbation expansion around
these backgrounds.

\Date{1/92}

\centerline{\it Introduction and General Discussion}

Many of the important questions in string theory circle around the issue
of background independence.  String theory, as a theory of quantum gravity,
should dynamically pick its own spacetime background.  We would expect
that this choice would be independent of the classical solution around
which the theory is initially  defined.   We may hope that the
theory finds a unique ground state which describes our world.

We can try to draw lessons that bear on these questions from the exactly
solvable matrix models of low dimensional string theory \mat \dbsc \cargese.
The physics of these models depends on a variety of parameters -- to
all orders in perturbation theory in every case, and in those models
that are well defined, nonperturbatively as well.  Example of these
parameters (which can be considered superselection sector labels) include
the KdV times $t_k$ in the $c<1$ systems and the cosmological constant
and radius of the $c=1$ system\ceqone.  One might think that the presence of
such superselection sectors is a special property, related in some way to the
integrability of the matrix models.  It will be one goal of this paper
to argue that this is not the case, and that, at least to all orders in
perturbation theory, it is a rather generic phenomenon.  Much of our
exposition is elementary but because these questions have been the
source of a certain amount of confusion both for ourselves and others we
thought it worthwhile to present a full discussion.  We try to provide
simple and intuitive pictures of some known results.
Our general
discussion applies both to field theory (where these phenomena are well
known) and to string theory.  We will present the arguments
in the language of string theory.

We will be discussing strings embedded in a Euclidean target space.  In
such a situation we often restrict attention to operators
in the world-sheet conformal field theory whose
dimensions are bounded from below.  These correspond to (delta-function)
normalizable wavefunctions in the conformal field theory Hilbert space.
In a space-time description the kinetic term
in a target space ``string field theory'' Lagrangian is
\eqn\kinterm{\Psi \CK \Psi }
where, essentially,
$\CK= L_0-1$.  Therefore the eigenfunctions of $\CK$ are the same as
the eigenfunctions of $L_0$. It is a basic principle of quantum
mechanics that in the space-time functional integral we sum
over (delta-function) normalizable eigenfunctions of $\CK$.  In a first
quantized formalism (in field theory or in string theory) this fact
follows from sewing amplitudes by inserting a complete set of
normalizable states in intermediate channels.  Therefore, the quantum
fluctuations and the states which flow in loops have normalizable wave
functions and can be expanded in the normalizable eigenfunctions of
$\CK$.

Clearly, for a well defined Euclidean
functional integral over the string field
$\Psi$ the spectrum of $\CK$ should not only be bounded from below but
there should be no negative eigenvalues.  Let us examine the different
possible normalizable eigenmodes of $\CK$:
\item{1.}
Positive eigenmodes of $\CK$ -- ``stable modes.''  The integral over
them is damped and leads to finite effects in perturbation theory.
\item{2.}
Negative eigenmodes of $\CK$ -- ``unstable modes.''  The integral over
them is divergent.  In a proper time description of the amplitudes,
there are exponential divergences.  From a space-time point of view,
these correspond to an instability of the system and render the
expansion inconsistent.  If the system has a non-compact
direction and the spectrum of $\CK$ is continuous, the existence of such
negative modes implies the existence of associated (dressed)
zero modes which are on-shell in Euclidean space.
\item{3.}
Zero eigenmodes of $\CK$ -- ``on-shell modes.''  The integral over
these modes depends on the details of the interaction terms in the
Lagrangian.  The most interesting case is when there is no potential for
these modes.  Then they should be treated carefully.
The moduli of the known vacua of the critical
string are of this kind.  The existence of such modes can lead to
mild infrared divergences (power law in proper time) in one loop
diagrams.  Their consequences at higher orders in perturbation
theory depends on their couplings and will be discussed below.

It is also important for
our purposes to examine {\it non-normalizable}
eigenfunctions of $\CK$ although such wavefunctions are typically ignored.
Allowing exponentially growing wavefunctions in Euclidean space,
one finds many new solutions of the linearized equations of motion
\eqn\lineq{\CK\Psi=0~~.}
{} From the world-sheet point of view they correspond to $(1,1)$
operators.  For example, one can construct operators like $e^{kX}$ out
of the free field $X$ corresponding to a non-compact space coordinate.
Their dimensions, $\Delta=-\half k^2$, are not bounded from below for
$k$ real.  Using such negative dimension operators, other operators in
the theory can be ``dressed'' and be turned into $(1,1)$ operators
$O_k$.  At least infinitesimally, one can add such on-shell operators
with small coefficients $t_k$ to the world-sheet Lagrangian and
construct new backgrounds.  Such backgrounds can be thought of as the
analytic continuations of oscillating Minkowski time dependent solutions
to Euclidean space.

It is central to the point we are trying to make that
such backgrounds do not fluctuate.
As we discussed above, only the normalizable
eigenmodes of $\CK$ are subject to quantum fluctuations.  The ``state''
corresponding to the operator $O_k$ has non-normalizable (not even
delta-function normalizable) wavefunction, is out of the conformal field
theory Hilbert space, and therefore does not propagate in loops.
Therefore, the corresponding background shifting parameter $t_k$ does
not fluctuate.  It labels a superselection sector and cannot be changed
to all orders in perturbation theory.  Clearly, the vacuum amplitude
depends on $t_k$ but because it does not fluctuate, it will not relax to
a minimum -- every value is allowed and corresponds to a different
background\foot{We should mention a more familiar example of
superselection sectors, distinct from the one discussed in the text.
It occurs both in field theory and in string
theory with more than two non-compact asymptotically flat directions.
Whenever there is a set of truly
degenerate ground states (whether related by
a symmetry or not), the Hilbert space breaks into separate Hilbert
spaces (superselection sectors) built on top of every such ground state.
Examples of such superselection sectors occur in the
critical string and are labeled by the moduli of the conformal field
theory.  It is important that when
the non-compact directions are compactified these moduli
fluctuate.  This distinguishes this situation from the one described in
the text.   We thank J.
Distler and D. Kutasov for discussions of these points.}.
Since the field $\Psi$ is arbitrarily large at infinity, it
is intuitively clear why it is impossible to change the value of $t_k$.
Regardless of how small the perturbation $t_k$ is, its effect at
infinity is arbitrarily large.

In the Minkowski time picture of the
non-compact coordinate, the parameter $t_k$ corresponds to an initial
condition labeling the classical solution.  Since Minkowskian evolution
is oscillatory, the initial conditions have effects even at
long time.

In the rest of this note we will discuss several examples of these
phenomena in low dimensional string theory.  A common characteristic
of all these examples is the presence of a string coupling constant that
varies rapidly in space.

\bigskip

\centerline
{\it Examples: Theories with rapidly varying coupling constant}

An important class of backgrounds have a non-compact dimension $\phi$
and a dilaton field expectation value linear in $\phi$.  The quadratic part of
the target space effective Lagrangian is
\eqn\edpre{\CL = e^{-Q\phi} \left( \half (\partial_\phi \Phi )^2 + (\Delta
-1) \Phi^2  \right)}
where $\Phi$ is a generic target space field associated with an operator
of dimension $\Delta$ in the conformal field theory of all the other
coordinates.  In general, there are many such fields, and if there are
some more non-compact dimensions, even a continuum of such fields.  In
this case we should also include a graviton field and a dynamical
dilaton field.  However, in order to keep the notation simple we
pretend that there is only one $\Phi$.  Clearly, the string coupling
$g_{st}= e^{\half Q \phi}$ depends on $\phi$.  The theory is weakly
coupled for $\phi \rightarrow -\infty$ and strongly coupled for
$\phi\rightarrow \infty$.  In order to apply our analysis above we
should redefine the fields such that the kinetic term is canonical as in
\kinterm.  Defining
\eqn\fieldred{\Psi=g_{st}^{-1} \Phi=  e^{- \half Q \phi} \Phi ~~,}
and integrating by parts in $\phi$, the kinetic term \edpre\ becomes
\eqn\wouted{ \half (\partial_\phi \Psi )^2 +\half m^2 \Psi^2 }
with
\eqn\msqu{\half m^2 = \Delta + {Q^2 \over 8} -1 }
and we recognize $m$ as the mass of the field $\Psi$.  The interaction
terms have roughly the form
\eqn\intterm{{1 \over 3} e^{-Q\phi} \Phi^3= {1 \over 3}
e^{\half Q \phi} \Psi^3 ~~.}

The target space objects $\Phi$ and $\Psi$ play important roles in
the first quantized formalism.  $\Psi$ is the wavefunction of the first
quantized state and $\Phi=e^{\half Q \phi} \Psi $ is related to the
corresponding vertex operator $V$.  The extra factor of the string coupling
$e^{\half Q \phi}$ can be understood from a change of variables between
the sphere and the cylinder \natiliou\ \joeliou.  The change of variable
is singular at the location of the operator and induces a delta function
contribution to the two curvature $R^{(2)}$ which is consistent with the
change of the Euler character between these two topologies.  This has
the effect of shifting the world-sheet action by $\half Q \phi$ or,
equivalently, multiplying the operator by $e^{\half Q \phi}$.  Because of
this factor, the identity operator in the world-sheet theory is
associated with the wavefunction $e^{-\half Q \phi}$ which is not
normalizable at $\phi
\rightarrow -\infty$ and is not in the Hilbert space of the conformal
field theory.  Since the $SL(2,C)$ invariant state is not in the Hilbert
space, many of the standard properties of a conformal field theory are
not satisfied.  In particular, the correspondence between states and
operators breaks down \natiliou\foot{ Pointing out an analogy to a
similar situation in mathematics G. Zuckerman has suggested the term
``non-amenable quantum field theories'' for these theories.}.

Without modifying the background the theory is ill defined.  There is a
region in the target space $\phi \rightarrow + \infty$ where the
interactions \intterm\ are infinitely strong.  A consistent background
should prevent the string from reaching this region.  There should be a
``wall'' at $\phi$ of order one.  Two such walls have been discussed in
the literature.  One possibility is to turn on an expectation value of
some field $\Phi$ which modifies the naive kinetic term
\eqn\kinn{\CK_0= - \half \partial_\phi^2 + \half m^2 }
to
\eqn\newk{\CK= - \half \partial_\phi^2 + \half m^2 + <\Phi> }
with $\lim_{\phi \rightarrow +\infty} <\Phi>= +\infty$.  Of course, one
should make sure that this modification is a solution of the equations
of motion.  An example of such a wall is the cosmological constant in
Liouville theory \pol \ctthrn \gervais \natiliou \joeliou.
Alternatively, since the target space metric is one of the fields, it
can be chosen so that the region of large positive $\phi$ is not in the
target space, as in the two dimensional Euclidean black hole \bh.  We
will discuss these two special cases in detail below.

We now examine the eigenmodes of the kinetic operator $\CK$ in this
situation.  The wall which suppresses the region of large $\phi$ affects
the functional form of the eigenfunctions.  The modes which fluctuate
are the (delta-function) normalizable eigenmodes.  With a soft wall as
in \newk\ the wavefunctions vanish rapidly as $\phi \rightarrow \infty$
and with a hard wall as in the black hole they should satisfy the proper
boundary conditions at the boundary of $\phi$.  In most interesting
cases the modification of the naive kinetic term \kinn\ is small as
$\phi \rightarrow -\infty $, and the analysis there is simple.  A
wavefunction with $\CK$ eigenvalue $\lambda> \half m^2$ behaves there as
$\sin k(\phi +\phi_0)$ with $k^2 +m^2=2\lambda$ and is delta function
normalizable.  A wavefunction with eigenvalue $\lambda< \half m^2$
behaves there as $e^{k(\phi +\phi_0)}$ with $k=\sqrt{m^2-2\lambda}$ and
has finite norm (``bound state'').  Depending on the details of the
theory there might or might not be negative $\lambda$ normalizable
eigenmodes.  If there are, the background is unstable.  Otherwise, all
the normalizable eigenmodes have positive $\lambda$ and the theory is
finite.  If there is a $\lambda=0$ eigenmode or if there is a continuum
$\lambda \in (0,\infty)$ a more careful analysis is necessary (see
below).

As we mentioned above, we should also consider the non-normalizable zero
modes of $\CK$.  These can be turned on as perturbations.  Since they do
not fluctuate, they label superselection sectors.  Just as for the
normalizable modes, to ensure a sensible perturbation theory we should
impose proper boundary conditions in the strong coupling region $\phi
\rightarrow \infty$ -- the wavefunction should decay.  Then the lack of
normalizability can come only from a divergence in the weak coupling
region $\phi \rightarrow -\infty$.

Unlike the standard backgrounds of the critical string ($Q=0$) here the
mass $m^2$ is not equal to $2(\Delta-1)$.  Since $\Delta$ is the
dimension of an operator in the conformal field theory, the standard
identification of relevant operators with tachyons, irrelevant ones with
massive and marginal operators with massless states breaks down
\natiliou.  In particular, truly marginal operators with $\Delta=1$ are
massive. The solution of the equation of motion $\CK \Psi=0$ depends on
$\phi$ and is not normalizable.  Hence, unlike the $Q=0$ case, here the
moduli do not fluctuate (again, to all orders in perturbation theory).

There is a simple spacetime picture of this lack of fluctuation.
Imagine doing the functional integral over $\Phi$ with Lagrangian
\edpre\ as weight.  In the region $\phi \rightarrow -\infty$ the
string coupling vanishes and so the quantum fluctuations of $\Phi$ are
strongly suppressed.  Any classical background set in this region will
remain locked.  We should emphasize that this locking is field
theoretic in character -- there is nothing intrinsically ``stringy''
about it.

A more familiar (and closely related) example of this
locking phenomenon occurs in field theories defined on a noncompact
base space of constant negative curvature \cw, e.g., the Poincar\'e
disk.  Here the constant zero eigenmode of the Laplacian is
exponentially non-normalizable because of the exponentially large amount
of volume at large distance. The normalizable spectrum is massive.  The
constant mode does not fluctuate, to all orders in perturbation theory,
and functions as a superselection parameter.  Here the locking is most
naturally thought of as arising from the huge volume at infinity rather
than the small coupling constant of the above discussion, but the
distinction is primarily semantic.  We should remind the reader that
nonperturbative effects in these theories can be much larger than in
flat space \cw\ and can in fact cause the zero mode to fluctuate.

\bigskip

\noindent
{\it 2D gravity:}

An important subclass of the theories with a rapidly varying coupling
constant are the ``non-critical strings.''  These can also be
interpreted as two-dimensional gravity coupled to matter \pol\plsltw.
The field $\phi$ is the conformal factor of the two-dimensional metric
\dewitt\phitime. They have a non-perturbative realization as matrix
models.  The wall is provided by the world sheet cosmological constant.
Thinking about these theories as theories of gravity provides us with a
complementary interpretation of the discussion above.

The insertion of a state into the functional integral is obtained by
cutting a little hole and gluing in the appropriate wavefunction.  The
length of the boundary in the fluctuating metric $\ell=\oint
e^{\gamma\phi/2}$ is one of the degrees of freedom in the wave function.
Surprisingly, it turns out that the Liouville parts of the wave functions
$\Psi(\phi = {2 \over \gamma} \log \ell)$ satisfy the minisuperspace
Wheeler-de-Witt equation \mss
\eqn\wdwmm{\left(-\half \partial_\phi^2+\half \gamma^2 \mu\ell^2
+{\gamma^2 \over 8} \nu^2 \right) \Psi_\nu (\phi)=0 }
which is the equation for an eigenfunction of the kinetic term \newk.
Demanding that the wave function suppresses infinite size holes leads to
the IR boundary condition $\lim_{\phi \rightarrow \infty}\Psi
(\phi)=0$ which implies
\eqn\wavfun{\Psi_\nu ={ 1\over \pi } \sqrt{\nu \sin (\pi \nu)}
K_\nu(2 \sqrt \mu \ell)~~.}
In the general discussion above this boundary condition was motivated by
suppressing the wave function in the region of infinitely strong
coupling
\foot{We would like to note in passing a peculiarity of two-dimensional
gravity with a negative cosmological constant.  Clearly, the Euclidean
functional integral is unstable.  However, this theory might be a
consistent
Minkowskian world-sheet theory without topology change so that the
tachyons are not excited.  Unlike the
positive cosmological constant theory, the gravitational wave functions
of all matter states are normalizable.  The two linearly independent
solutions of \wdwmm\ behave like $\ell^{-\half} e^{\pm 2 i
\sqrt{|\mu|}\ell}$ as $\phi \rightarrow +\infty$.  Any linear
combination of them is normalizable there.  This freedom can be used to
pick a normalizable solution as $\phi \rightarrow -\infty$ even for real
$\nu$ in \wdwmm.}.

The delta function normalizable wave functions with imaginary $\nu$
propagate in loops and those with real $\nu$ diverge in the UV ($\phi
\rightarrow -\infty$) and do
not fluctuate.  The latter label different backgrounds for the string.
The gravitational interpretation of this fact \natiliou\ is as follows.
The states which propagate in handles have finite size.  Hence their
wave functions are localized in $\phi $ space and are spanned by the
delta function normalizable states with imaginary $\nu$.  Since these
states have finite size they are called macroscopic.  The states
associated with real $\nu$ are not normalizable.  Their wave functions
are peaked at short distance and therefore they describe microscopic
holes.  Clearly only they can be associated with local operators
\natiliou.  In this context world-sheet locality provides a natural
reason to examine non-normalizable states.

In the general discussion above we motivated non-normalizable
deformations by comparing them with an analytic continuation of
backgrounds which oscillate in Minkowski time.  This Euclidean time
picture is also natural in gravity.  Here we can define the theory at
some distance scale \dewitt\ $\phi$ and let it evolve by the
renormalization group into the IR.  In this picture, as many workers
have pointed out, $\phi$ plays the role of the renormalization group
time.  The values of the fields at some scale are the initial conditions
for this renormalization group evolution and they affect the final
answer.  Therefore, one should not average over them but study the
answers as a function of these initial conditions.  This is the
two-dimensional gravity version of the background dependence and
superselection sectors we discussed above.

\bigskip

\noindent
{\it The $c=1$ model:}

All the physical operators in the $c<1$ minimal models are dressed with
real $\nu >0$ Liouville wave functions and are massive.  The $c=1$
system has a massless operator -- the identity operator.  Its on-shell
wave function $K_0(2 \sqrt \mu \ell) $ is proportional to $\phi$ at
$\phi \rightarrow -\infty$.  Hence it is not normalizable and the local
vertex
operator is $\phi e^{{Q \over 2} \phi}=\phi e^{\gamma \phi}$ rather than
$e^{\gamma \phi}$ \joecone\ (the latter decouples \kutdf).  By the
general discussion above, its coefficient in the action, $\mu$, does not
fluctuate and corresponds to a superselection sector.

The cosmological constant is almost a
macroscopic state; it is at the bottom of the continuum of normalizable
states with wave functions $\Psi_\nu ={ 1\over\pi } \sqrt{\nu \sin (\pi
\nu)} K_\nu(2 \sqrt \mu \ell)$ with $\nu$ imaginary.  The fact that it
is not macroscopic -- it cannot quite propagate in intermediate channels
-- should have a signature in correlation functions.  To examine this in
more detail, consider the four point function of four microscopic states
with $X$ momenta $p_1, p_2, p_3, p_4$.  Since $\nu$ is not conserved,
the singular part of the amplitude is given by
\eqn\fouram{\int_0^{i\infty} d\nu {f_{p_1, p_2}(\nu)f_{p_3, p_4}(\nu)
\over \nu^2 - (p_1+p_2)^2  }   }
with the integral along the imaginary $\nu$ axis \natiliou.  The
function $f_{p_1, p_2}(\nu)$ is a form factor for the on-shell
microscopic tachyons of momenta $p_1,p_2$ to couple to the macroscopic
state with imaginary $\nu$.  It can also be thought of as an operator
product coefficient (except that there is no local operator with
imaginary $\nu$).  The expression for the four point function
\joeclas\moore\kutdf\ shows that the singularity of the amplitude as
$p_1+p_2 \rightarrow 0$ is proportional to $|p_1+p_2|$.  This translates
into $f(\nu ) \sim \nu$ as $\nu \rightarrow 0$ thus exhibiting
the decoupling of the vanishing $\nu$ state.  We therefore expect that a
general vertex of three macroscopic states $f(\nu_1, \nu_2, \nu_3) \sim
\nu_i$ as any $\nu_i \rightarrow 0$.

The following simple
heuristic argument explains this result.
For small imaginary $\nu_i$ we expect the form factor to be
given by
\eqn\formfa{f(\nu_1, \nu_2, \nu_3)=<\Psi_{\nu_1}|\Phi_{\nu_2}|\Psi_{\nu_3}>
= \int d\phi e^{{Q\over 2} \phi}
\Psi_{\nu_1}(\phi) \Psi_{\nu_2}(\phi) \Psi_{\nu_3} (\phi) ~~.}
The wall causes $\Psi$ to decay quickly as $\phi \rightarrow \infty$ and
$g_{st}=e^{{ Q\over 2}\phi}$ falls rapidly as $\phi \rightarrow -\infty$
so the integral in \formfa\ is dominated by a finite region around $\phi
= 0$.  For $\phi$ finite and $\nu \rightarrow 0$, the wave function
$\Psi_{\nu}(\phi)=N_{\nu}\Psi_{0}(\phi)$ where $N_{\nu}$ is a
normalization factor.  For $\nu$ fixed as $\phi \rightarrow -\infty$,
$\Psi_{\nu}(\phi)$ must become a linear combination of solutions of the
free WdW equation, $\sin (\nu \phi)$ and $\cos (\nu \phi)$.  We demand
smooth matching and delta function normalizability.  Using the fact
that $\Psi_0$ has nonvanishing derivative in the matching region
($\Psi_0 \sim \phi$ as $\phi \rightarrow -\infty$) we
see that $\sin (\nu \phi)$ is the dominant solution and that $N_{\nu}
\sim \nu$.  For small $\nu$ the wavefunction is small near the wall.
(Of course, one can derive this result from the known
properties of the Bessel functions.)  This is a signature of the $\nu=0$
state not being normalizable.  Therefore, as any $\nu_i$ approaches
zero, the form factor is proportional to $\nu_i$.  If all $\nu_i$ are
small,
\eqn\formtwo{f(\nu_1, \nu_2, \nu_3) \sim \nu_1 \nu_2 \nu_3 ~~.}
We have thus recovered the familiar derivative coupling of the tachyon
field \dj \moore\ \joeclas \kutdf \wadia \ms \mooreplesser.

This low $\nu$ behavior is important for the finiteness of the theory.
Divergences in string theory are IR divergences associated with on-shell
states in intermediate lines.  Since the on-shell zero momentum tachyon
is a massless state and is almost normalizable, there could be such
IR divergences.  More explicitly, the sum over intermediate states
involves integrals similar to \fouram\ along the imaginary $\nu$ axis
which could diverge from the region near $\nu=0$.  The form factor $f$
suppresses this region and renders these integrals finite\foot{For
finiteness we really need to know that the form factor for one tachyon
with any two other states $\sim  \nu$ for small $\nu$. This follows by
a very similar argument.}.

As an aside let us make a few remarks about the  high momentum behavior
of the amplitudes.  In a beautiful
calculation Moore \moore\ has shown that the high momentum behavior
of the $c=1$ theory is dominated by non-perturbative effects.  There is
a signal of this in the high momentum behavior of the fixed genus
amplitude -- it grows like a power of the momentum.  This
can be explained qualitatively as follows\foot{This point was understood
in discussions with T. Banks, D. Kutasov and G. Moore.}.  The tachyon
background\foot{The minisuperspace wave functions in the $c=1$ system
satisfy the Wheeler-de-Witt equation \wdwmm\ with $<\Phi> =\mu e^{\gamma
\phi}$ \moore\mss\ rather than with $<\Phi> =\mu \phi e^{\gamma \phi}$.
We do not understand how this is consistent with the fact that the
cosmological constant operator is $\phi e^{\gamma \phi}$.} in \newk\ is
needed in order to prevent the theory from sliding into the strong
coupling region $\phi\rightarrow \infty$.  However, since this wall is
soft, the more energy the system has the deeper it can probe into the
strong coupling region and the more important the quantum effects are.
To make this picture somewhat more quantitative, consider the two point
function of tachyons with momentum $p$ and study it in the semiclassical
approximation.  Following the semiclassical analysis in \natiliou, the
amplitude is dominated by a constant positive curvature surface
representing the reflection off the wall.  The classical motion in the
Liouville potential has energy $p^2$, and therefore the turning point of
the classical trajectory $\phi_t$ is determined by $\mu e^{\gamma
\phi_t}= p^2$.  The string coupling at that point is $g_{st} =g_0
e^{\half Q\phi_t}=g_0 {p^2 \over \mu}$.  We therefore expect the
effective string coupling at high momentum to behave like ${p^2\over
\mu}$.  This is precisely the behavior found in \moore\foot{Note
that this argument gives the conceptual basis for the difference between
the high momentum behavior in the $c=1$ model and in the $D=26$
bosonic string\grossmende.}.  Note that the
string coupling gets strong in the space-time UV and in world-sheet IR.
Such an inverse relationship between
world-sheet and space-time scales
has been noted in critical string theory \lenny.

Let us now return to our main theme.
Typically the linearized equation of motion \lineq\ does not have
normalizable solutions.  There are however some familiar examples of
such zero modes of $\CK$ in the critical string.  The zero momentum mode
of every massless field is a (delta-function) normalizable zero mode of
$\CK$.  Its wave function is independent of the non-compact directions.
In the situation with a $\phi$ dependent coupling constant we
impose the boundary condition that the wave function vanishes in the
strong coupling region $\phi \rightarrow
\infty$.  With such a boundary condition at one end it is non-generic
to find solutions of \lineq\ which do not diverge at the
other end, $\phi \rightarrow -\infty$.  There is however a notable
example of such a ``miraculous solution.''  It occurs in the Ramond
sector of the ``non-critical'' fermionic string.
The supersymmetric matter ground state has
$\Delta= {\hat c \over 16}$ and is always massless.  Unlike the
non-supersymmetric minimal models which have only massive operators,
half of the supersymmetric minimal models have such $\Delta= {\hat c
\over 16}$ states which are massless.  Furthermore, the minisuperspace
Wheeler-de-Witt equation is
\eqn\wdwmms{\left(-\half \partial_\phi^2+ \half |W'(\phi)|^2 - \half
W''(\phi) +{\gamma^2 \over 8} \nu^2 \right) \Psi_\nu (\phi)=0 }
where the superpotential $W$ is
\eqn\suppot{ W(\phi) = \mu e^{\gamma \phi}  ~~. }
This problem was first analyzed in \curt.
The $\nu=0$ eigenfunction is, as always in supersymmetric theories,
\eqn\nuzerosu{\Psi_0= e^{-W} = e^{-\mu e^{\gamma \phi} } ~~.}
It
satisfies the boundary conditions in the strong coupling region $\phi
\rightarrow \infty$, and approaches a constant in the weak coupling
region $\phi \rightarrow -\infty$.  Note that unlike the zero momentum
tachyon in the $c=1$ system which is also massless, the wave function of
this state is delta function normalizable.  Hence, the corresponding
state is macroscopic and it fluctuates.  It is not clear to us whether
this state decouples (as similar states decouple in the critical
string), or if not, whether it leads to IR divergences in perturbation
theory.

\bigskip

\noindent
{\it 2D Euclidean black hole:}

As another example of background dependence, we now discuss the 2D
Euclidean
black hole \bh.  The world-sheet Lagrangian is characterized by the
cigar metric and dilaton
\eqn\bhmet{\eqalign{
ds^2=& dr^2 +\tanh ^2r d\theta^2 \cr
D=& D_0 + \log \cosh ^2 r  ~~. \cr } }
The comparison with the $c=1$ system is obtained by writing
\eqn\bhconec{ \phi = - {1\over Q}(\log M + \log \cosh ^2 r) ~~}
in terms of the parameter $M$ which plays the role of the mass of the
black hole.  Here $Q=2\sqrt 2$ corresponds to $k={9 \over 4}$ in the
coset construction.  In these coordinates the dilaton is linear in
$\phi$ and the metric is
\eqn\bhmetn{ds^2={Q^2\over 4(1- Me^{Q\phi} )} d\phi^2 +(1- M e^{Q\phi} )
d\theta^2 ~~. }
Note that unlike the $c=1$ system, here $\phi$ has a maximum allowed
value, $-{1\over Q}\log M$, which is obtained at the tip of the cigar,
$r=0$.  Instead of the cosmological constant wall, here the region of
strong coupling is not accessible at all.  In the weak coupling region,
$\phi \rightarrow -\infty$ ($r\rightarrow \infty$) the metric is
approximately
\eqn\apprmet{ds^2 \approx {Q^2 \over 4}(1+Me^{Q\phi} )d\phi^2
+ (1 - M e^{Q\phi}) d\theta^2 ~~.}
The first signal of the wall as we approach it from this region is the
existence of the world-sheet operator $e^{Q\phi}({Q^2\over 4}
\partial \phi\bar \partial \phi - \partial\theta \bar
\partial \theta)$.  This operator is $(1,1) $ because $ \partial\theta
\bar \partial \theta$ and $ \partial\phi \bar
\partial \phi $ are $(1,1)$ and $e^{Q\phi}$ is $(0,0)$.  However, it
seems to violate the bound on the exponent $\alpha < {Q\over 2}$
\natiliou.  Furthermore, one might think that since $\alpha=Q$,
the corresponding wave function of this operator behaves like $e^{\half
Q \phi }$ as $\phi \rightarrow -\infty$ and hence it might be
normalizable.  If so, the mass of the black hole $M$ could fluctuate.

In fact the situation is more subtle.  It is misleading to view this
system as a perturbation of the problem without the wall.  The topology
of the non-zero $M$ target space is different from flat space with
$M=0$.  We should define the mass operator as the operator which induces
infinitesimal changes of $M$ around non-zero $M$.  It is obtained by
varying the world-sheet action with respect to $M$
\eqn\varym{V_M=\int
{Q^2 e^{Q\phi}  \over 4(1- Me^{Q\phi} )^2} \partial \phi \bar \partial \phi
-  e^{Q\phi} \partial \theta \bar \partial \theta }
whose wave function seems normalizable at $\phi \rightarrow -\infty$.
However, a properly computed norm of this metric deformation
\eqn\normmet{ |\delta g|^2= \half
\int\sqrt g  (g^{\alpha\gamma} g^{\beta \delta} +
g^{\alpha\delta} g^{\beta \gamma})
\delta g_{\alpha \beta} \delta g_{\gamma \delta}   }
diverges at the tip of the cigar.  Hence, this deformation is not
normalizable and the parameter $M$ cannot fluctuate\foot{A similar
calculation leads to a similar conclusion in  four dimensions \gpy. Other
(unstable) modes are important there, though\hawk \gpy.
We thank M. Douglas for pointing this out to us.}.  Equivalently, a
shift of $M$ is equivalent to a shift of $D_0$ which determines the
string coupling at the tip of the cigar.  The operator corresponding to
such a deformation is $\partial \bar \partial \phi$ whose wave function
is proportional to $e^{-\half Q\phi}$.  It is not normalizable as $\phi
\rightarrow -\infty$.  Hence the string coupling and $D_0$ (which is
related to the mass $M$) cannot fluctuate.  We conclude that there is a
one parameter family of backgrounds labeled by $M$ or $D_0$ which do not
explore each other to all orders in perturbation theory.

To further examine the stability of the black hole and the finiteness of
the perturbation expansion, we should look for negative or zero modes of
the small fluctuation operator $\CK$ which are normalizable.  We
restrict our attention to the $Q=2 \sqrt 2$, $k={9 \over 4}$ case.
The genus
one calculation of \gk\ shows that there are no localized negative modes
(which would have led to instabilities) and that the entire light
spectrum comes from the massless ``tachyon'' field
\foot{But note that the alternate quantization of \marsha\ predicts an
instability for $k<3$.}.  First we examine
the zero momentum tachyon.  Its wavefunction is \dvv\gk
\eqn\bhtac{\Psi_{0}(r)=  {1 \over g_{st}(r)}
\int d \theta (\cosh 2r + \cos \theta
\sinh 2r)^{-\half} ~~. }
It is proportional to $r$ (the generic behavior)
as $r \rightarrow \infty $ and is not normalizable.

Repeating the analysis of the form factor in the $c=1$ system, it is
easy to show that form factors here also vanish like the $r$ momentum,
$\nu$, as
$\nu\rightarrow 0$. Again this is a signature of the non-normalizability
of the zero momentum state.  Therefore, even though there are
normalizable eigenmodes of $\CK$ with arbitrarily small eigenvalues,
these modes decouple at low momentum and there will be no IR divergences
in the genus expansion. We thus argue that the vacuum amplitude genus
expansion of the 2D Euclidean black hole is finite to all orders in the
genus expansion (except for a trivial volume divergence at genus one).

The implications of these results for the Minkowski space
evolution of the black hole are not clear to us\foot{See \cern\ for a
discussion of related issues.}.

The nature of the wall in this theory is different from that of the
$c=1$ theory.  It is a hard wall completely preventing the string from
reaching the strong coupling region.  Therefore, we speculate that the
high momentum behavior of the amplitudes is softer here than in the
$c=1$ system.  The effective coupling constant as a function of the
momentum either does not grow indefinitely, or grows slower than in the
$c=1$ theory.

In summary, we have argued that the presence of the kind of
superselection parameters observed in the matrix models is rather
generic, at least to all orders in perturbation theory.  It is due to
the presence of non-normalizable deformations, a known phenomenon in
field theory.  The persistence of such behavior nonperturbatively is an
interesting and for the most part open question.  Since such
superselection sectors occur in the critical string, we should
understand why our world is not ``stuck'' in such a state and is rather
in its known background with four non-compact flat dimensions.  In
particular, why is the world not in a ``non-critical'' string
background?

{\it Note added.} E. Witten has also argued
for the stability of the 2D black hole.

\vfill\eject

\centerline{\bf Acknowledgements}

It is a pleasure to thank T. Banks, J. Distler, M. Douglas, D. Friedan,
K. Gawedzki, A. Kupiainen, E. Martinec, G. Moore, L. Susskind, E.
Witten, A.B.  Zamolodchikov and especially D.  Kutasov for useful
discussions.  This work was supported in part by DOE grant
DE-FG05-90ER40559.

\listrefs
\bye